\documentclass[10pt,journal]{IEEEtranTCOM}

\usepackage[english]{babel}
\usepackage[usenames]{color}
\usepackage[cp1250]{inputenc}
\usepackage{amsfonts}
\usepackage{amsthm}
\usepackage{graphicx}
\usepackage{epsfig}
\usepackage{mathrsfs}
\usepackage{amsmath}
\usepackage{algorithm}
\usepackage{algorithmic}
\usepackage{hyperref}
\DeclareMathOperator*{\argmax}{arg\,max}

\theoremstyle{plain}

\begin{document}
\newcommand{\bea}{\begin{eqnarray}}
\newcommand{\eea}{\end{eqnarray}}
\newcommand{\be}{\begin{equation}}
\newcommand{\ee}{\end{equation}}
\newcommand{\beas}{\begin{eqnarray*}}
\newcommand{\eeas}{\end{eqnarray*}}
\newcommand{\bs}{\backslash}
\newcommand{\bc}{\begin{center}}
\newcommand{\ec}{\end{center}}
\def\SC {\mathscr{C}}

\title{Adaptive Student's t-distribution with\\ method of moments moving estimator \\ for nonstationary time series}
\author{\IEEEauthorblockN{Jarek Duda}\\
\IEEEauthorblockA{Jagiellonian University, Institute of Computer science and Computational Mathematics,\\ \L{}ojasiewicza 6,
 Krakow, Poland,
Email: \emph{dudajar@gmail.com}}}
\maketitle

\begin{abstract}
The real life time series are usually nonstationary, bringing a difficult question of model adaptation. Classical approaches like ARMA-ARCH assume arbitrary type of dependence. To avoid their bias, we will focus on recently proposed agnostic philosophy of moving estimator: in time $t$ finding parameters optimizing e.g.  $F_t=\sum_{\tau<t} (1-\eta)^{t-\tau} \ln(\rho_\theta (x_\tau))$ moving log-likelihood, evolving in time. It allows for example to estimate parameters using inexpensive exponential moving averages (EMA), like absolute central moments $m_p=E[|x-\mu|^p]$ evolving for one or multiple powers $p\in\mathbb{R}^+$ using $m_{p,t+1} = m_{p,t} + \eta (|x_t-\mu_t|^p-m_{p,t})$. Application of such general adaptive methods of moments will be presented on Student's t-distribution, popular especially in economical applications, here applied to log-returns of DJIA companies. While standard ARMA-ARCH approaches provide evolution of $\mu$ and $\sigma$, here we also get evolution of $\nu$ describing $\rho(x)\sim |x|^{-\nu-1}$ tail shape,  probability of extreme events - which might turn out catastrophic, destabilizing the market.
\end{abstract}
\section{Introduction}
\begin{figure}[t!]
    \centering
        \includegraphics[width=8.5cm]{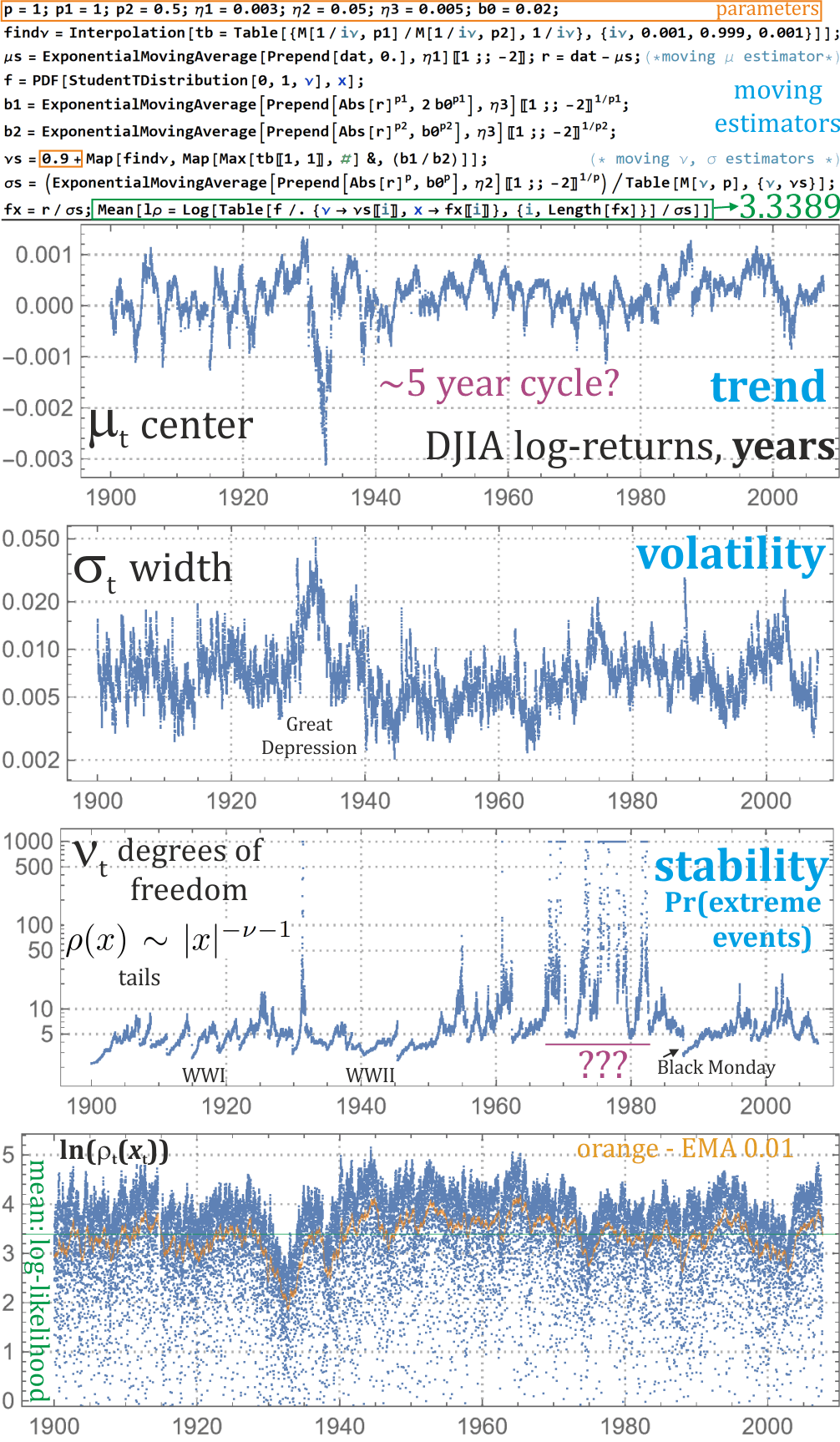}
        \caption{Mathematica code used for moving estimation of all $\theta=(\mu,\sigma,\nu)$ Student's t-distributions parameters (using $M_{\nu p}=E[|(x-\mu)/\sigma|^p]$ moment formula (\ref{mom})), and results of its application to 107 years of daily log-returns of DJIA (Dow Jones Industrial Average) time series. The parameters were manually tuned for this case to maximize log-likelihood: mean $\ln(\rho_t(x_t))$ showed at the bottom. We can see interesting evolution through this century which might be worth a deeper investigation, like $\approx 5$ year period cyclic behavior of the center $\mu$, huge $\approx 25\times$ change of width $\sigma$, and a few nearly Gaussian $\nu\to \infty$ periods mostly during 1967-1983. While $\mu$ describes the general up/down trend, $\sigma$ is close to volatility, additional $\nu$ complements it with kind of stability - probability of potentially catastrophic extreme events.}
       \label{djiastud}
\end{figure}

Choosing a parametric family of probability distributions, e.g. Student's t-distribution here, there is usually focus on intuitively \textbf{static estimation}: optimization of a single set of parameters $\theta$ for the entire dataset, usually through maximization of some evaluation like $F=\frac{1}{T}\sum_{t=1}^T f(\theta,x_t)$. For example log-likelihood in popular MLE (maximal likelihood estimation) using $f(\theta,x)=\ln(\rho_{\theta}(x))$, where $\rho_{\theta}(x)$ is PDF (probability distribution function) for the assumed parametric family. This way all datapoints have equal $1/T$ contributions, what seems a perfect choice for stationary time series.

\begin{figure*}[t!]
    \centering
        \includegraphics[width=18cm]{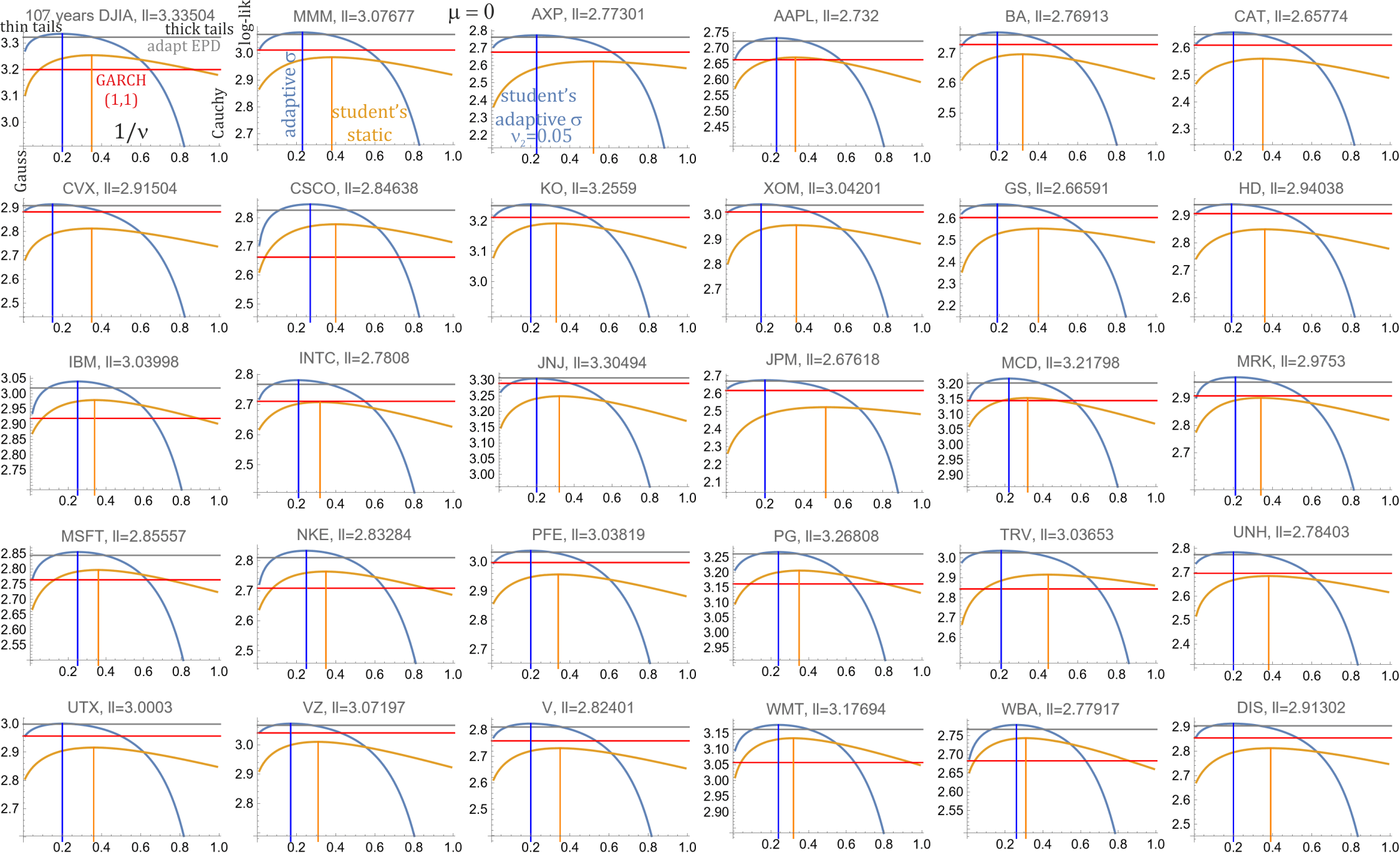}
        \caption{Log-likelihoods (mean $\ln(\rho_t(x_t))$) evaluations for log-returns of 107 years DJIA time series, and 10 years for 29 individual companies. In horizontal axis there is $1/\nu$ Student's t-distribution degrees of freedom (from Gauss to Cauchy distributions), for static parameters (orange), and adaptive $\sigma$ scale parameter (blue, using $p=1$ power and $\eta_2=0.05$ learning rate), all for $\mu=0$ center. We can see adaptation has allowed for less heavy tails (larger $\nu$ in maximum). There are also shown analogously the best from $\sigma$ adaptation for $\rho(x)\sim \exp(-|x|^\kappa)$ Exponential Power Distribution in the previous article~\cite{adaptive} (gray). Red line shows evaluation of $\sigma$ adaptation by standard GARCH(1,1) model - which is comparable with $\nu=\infty$ Gaussian case, but usually slightly worse.  }
       \label{eval}
\end{figure*}

In contrast, real life time series are often non-stationary, suggesting to use \textbf{adaptive estimation}~\cite{adaptive} instead - with evolving parameters, like $\theta_t=(\mu_t,\sigma_t,\nu_t)$ in Fig. \ref{djiastud} for  Student's t-distribution we will focus on. Moving estimator for each time $t$ will separately optimize $\theta_t$ parameters based on the previous values $\{x_\tau\}_{\tau<t}$ with weakening weights, to finally optimize:
\be F = \frac{1}{T}\sum_{t=1}^T f(\theta_t,x_t) \quad \textrm{e.g. log-likelihood: }\frac{1}{T}\sum_{t=1}^T \ln(\rho_{\theta_t}(x_t))\ee
A natural approach to estimate $\theta_t$ is optimizing analogous function $F_t$: using only the past values $\{x_\tau\}_{\tau<t}$, with exponentially weakening weights to get local behavior:
\be \theta_t=\argmax_\theta F_t\qquad\textrm{for}\qquad F_t = \sum_{\tau<t} \bar{\eta}^{t-\tau} \ln(\rho_{\theta}(x_\tau))\label{opt}\ee
for $\bar{\eta}\in (0,1)$ learning rate usually above 0.9, also define $\eta=1-\bar{\eta}$ for convenient calculation.

The above (\ref{opt}) moving MLE can be easily directly optimized for $\sigma$ scale parameter of EPD (exponential power distribution) $\rho(x)\sim \exp(-|x|^\kappa)$~\cite{adaptive} thin tail family containing e.g. Gauss and Laplace distributions, from \textbf{absolute central moments}: $m_p= E[|x-\mu|^p]$, for adaptation evolving with EMA (exponential moving average):
\be m_{p,t+1} = m_{p,t} + \eta (|x_t-\mu_t|^p-m_{p,t}) \ee
using $p=\kappa$ for EPD, and $\mu_t$ as constant or also adapted using EMA. Here we will take it to Student's t-distribution, this time not through direct MLE due to lack of explicit formula, but through method of moments instead - estimating $\sigma$ scale parameter from absolute central moment for a single power, or $\nu$ degrees of freedom from such moments for two powers.

On example of 107 years Dow Jones Industrial Average (DJIA) daily log-returns and 10 years for 29 its recent companies, there was tested such adaptive estimation especially of $\sigma$, leading to essentially better log-likelihood evaluation, here for Student's t-distribution slightly better than for EPD~\cite{adaptive}. Also essentially better than standard methods of $\sigma$ prediction like GARCH(1,1)~\cite{garch} - from one side focused on Gaussian distribution, but also arbitrarily assumed dependencies - here replaced with agnostic philosophy of moving estimator optimizing local parameters.

Such adaptive estimation can be combined with other methods, which might be added in later versions of this article.  For example online PCA~\cite{OPCA} or adaptive linear regression~\cite{regression} to combine information from multiple sources like companies here or macroeconomical data - e.g. to improve prediction of the moments, used for parameter estimation here. Finally, as discussed in \cite{adaptive}, we can use such parametric distributions for normalization $y_t = \textrm{CDF}_t(x_t)$, and then model slight distortion from uniform distribution of $\{y_t\}$ with HCR (hierarchical correlation reconstruction)~\cite{hcr} modelling density as a linear combination, in static or adaptive (evolving in time) way.

\section{Time series used for evaluation}
There was used 1900-2007 daily Dow Jones index\footnote{Source of DJIA time series: http://www.idvbook.com/teaching-aid/data-sets/the-dow-jones-industrial-average-data-set/},  working on $x_t=\ln(v_{t+1}/ v_t)$ sequence of daily log-returns.

Figure \ref{eval} additionally contains such evaluation of log-returns for 29 out of 30 companies used for this index in September 2018. Daily prices for the last 10 years were downloaded from NASDAQ webpage (www.nasdaq.com) for all but DowDuPont (DWDP) - there were used daily close values for 2008-08-14 to 2018-08-14 period ($2518$ values) for the remaining 29 companies: 3M (MMM), American Express (AXP), Apple (AAPL), Boeing (BA), Caterpillar (CAT), Chevron (CVX), Cisco Systems (CSCO), Coca-Cola (KO), ExxonMobil (XOM), Goldman Sachs (GS), The Home Depot (HD), IBM (IBM), Intel (INTC), Johnson\&Johnson (JNJ), JPMorgan Chase (JPM), McDonald's (MCD), Merck\&Company (MRK), Microsoft (MSFT), Nike (NKE), Pfizer (PFE), Procter\&Gampble (PG), Travelers (TRV), UnitedHealth Group (UNH), United Technologies (UTX), Verizon (VZ), Visa (V), Walmart (WMT), Walgreens Boots Alliance (WBA) and Walt Disney (DIS).

\begin{figure}[t!]
    \centering
        \includegraphics[width=9cm]{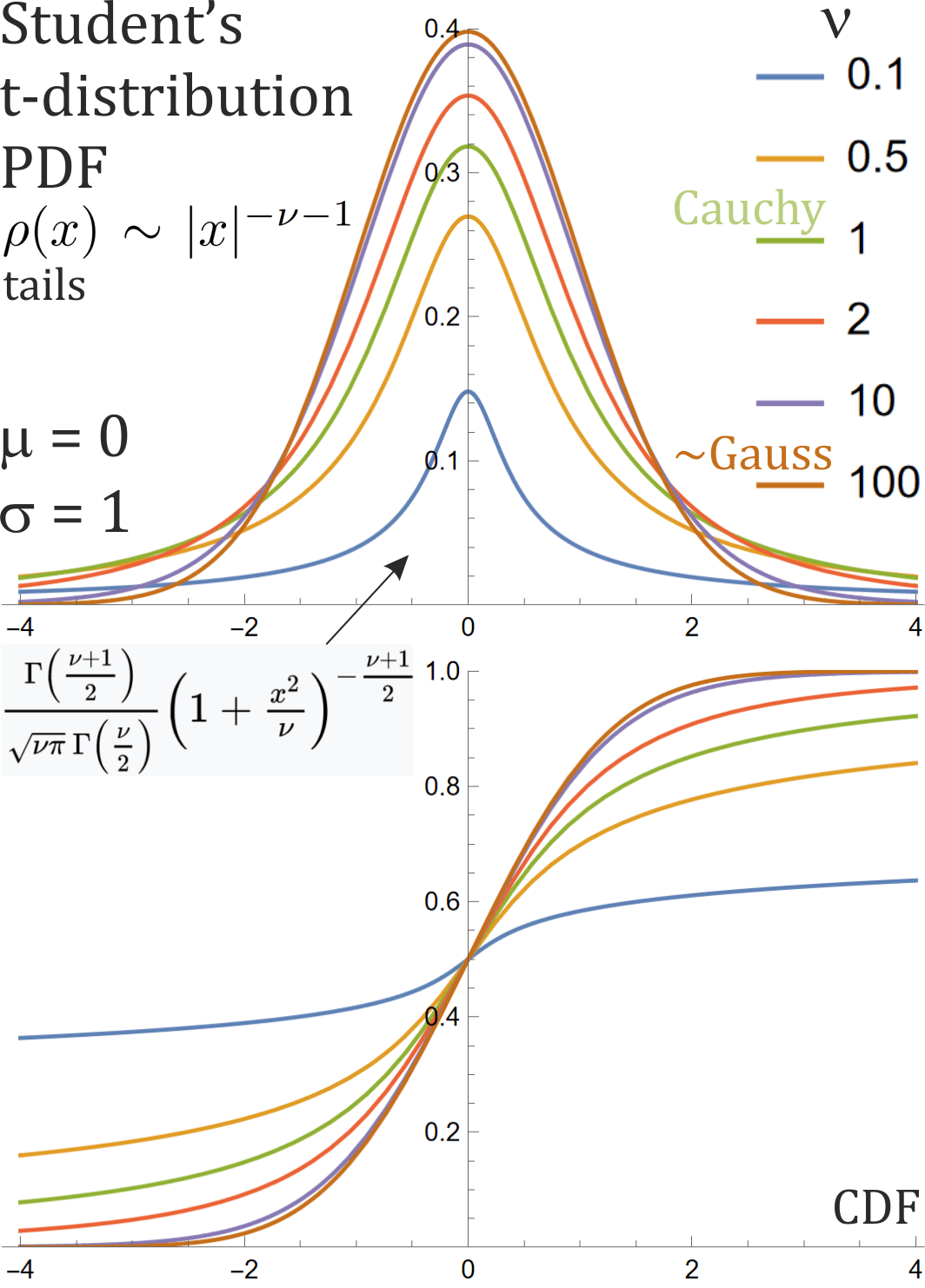}
        \caption{Probability distribution function (PDF, asymptotically $\sim |x|^{-1-\nu}$) and cumulative distribution function (CDF) for Student's t-distribution with fixed center $\mu=0$ and scale parameter $\sigma=1$, but various shape parameter $\nu$. We get Gaussian distribution for $\nu\to\infty$, Cauchy distribution for $\nu=1$, and can also cover different types of heavy tails and bodies of distribution.}
       \label{student}
\end{figure}

\begin{figure}[t!]
    \centering
        \includegraphics[width=9cm]{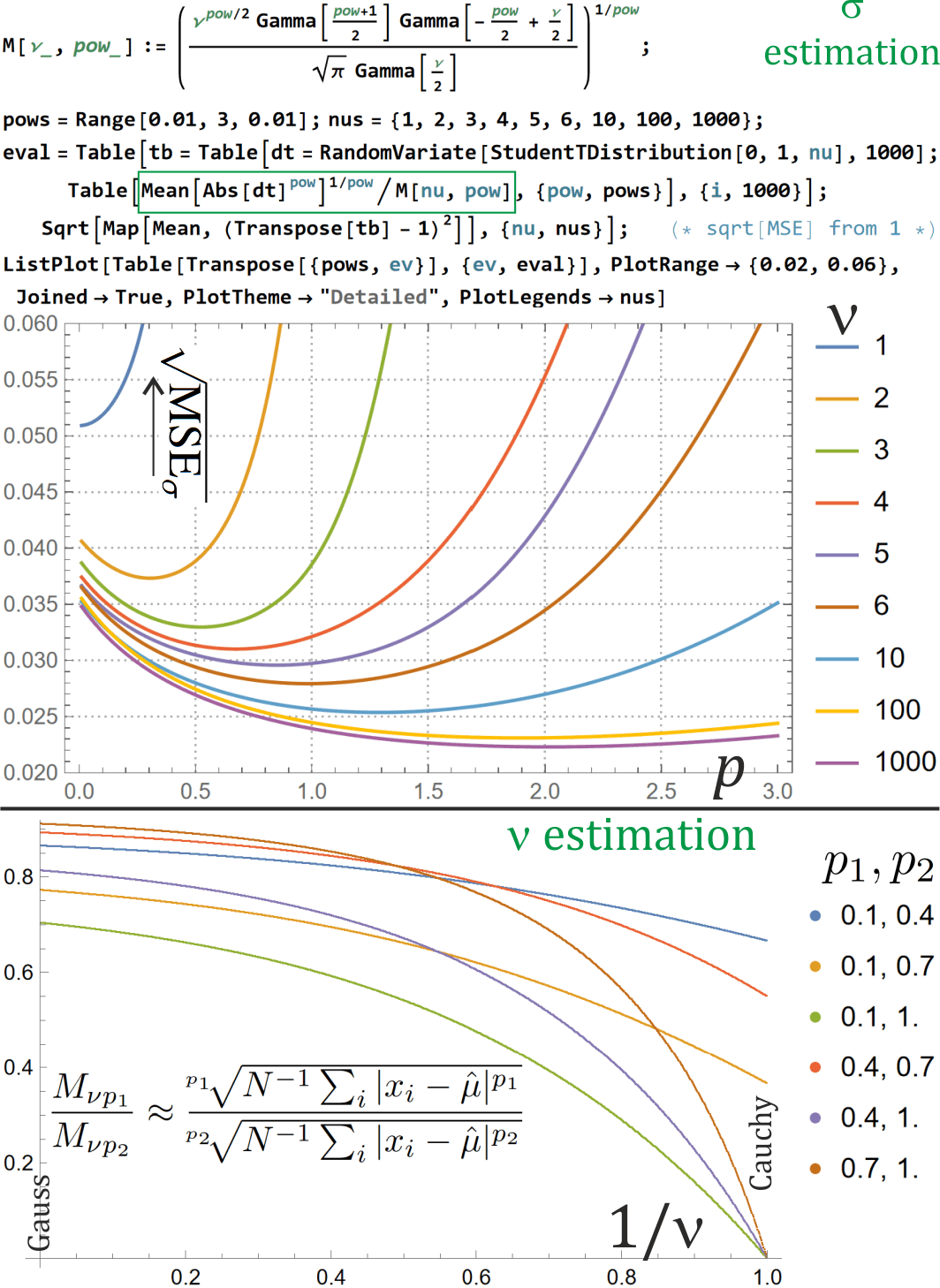}
        \caption{\textbf{Top}: error dependence for choice of power $p$ in $\sigma$ estimation as $\hat{\sigma} =  \sqrt[\leftroot{5}p]{T^{-1} \sum_t |x_i-\hat{\mu}|^p}/M_{\nu p}$. We can see that for Gaussian distribution $\nu\to\infty$ we should choose $p=2$ as in standard variance estimation, but to improve prediction should reduce this $p$ for lower $\nu$ to $p\approx \nu/6$. \textbf{Bottom}: monotonous functions for $\nu$ estimation for various choices of 2 powers $p_1,p_2$.  }
       \label{est}
\end{figure}

\section{Student's t-distribution and adaptation}
The Student's t-distribution was first introduced by Friedrich Helmert in 1875~\cite{student}, and later in 1908 by William Sealy Gosset signed as "Student"~\cite{student1}, leading to the popular name.

Its basic application is for distribution of sum of $\nu+1$ i.i.d. Gaussian random variables: for the difference between the sample mean and the real mean. For $\nu=1$ it is Cauchy distribution, for large $\nu\to\infty$ it approaches Gaussian distribution.

Its PDF (probability density function), shown in Fig. \ref{student}, is:
\be \rho_{\mu \sigma \nu}(x)\equiv \rho(x) =
\frac{\Gamma((\nu+1)/2)}{\sqrt{\nu\pi}\, \Gamma(\nu/2)}
\left(1+\frac{(x-\mu)^2}{\sigma^2 \nu}  \right)^{-\frac{1+\nu}{2}}
\ee
for $\mu\in\mathbb{R}$ and $\sigma,\nu\in \mathbb{R}^+$, $\Gamma(z)=\int_0^\infty t^{z-1} e^{-t} dt$ gamma function. Crucially, it has one over polynomial heavy tails $\rho(x)\sim |x|^{-\nu-1}$ for $|x|\to \infty$, hence finite moments $E[x^p]$ only for $p < \nu$.

Its CDF (cumulative distribution function) for $\mu=0, \sigma=1$ is below, for the general case substitute $x \to (x-\mu)/\sigma$:
\be \int_{-\infty}^{x} \rho_{01\nu}(y)dy=\frac{1}{2} +x\Gamma\left(\frac{\nu+1}{2}\right)
\frac{F_{1,2}\left(\frac{1}{2},\frac{\nu+1}{2};\frac{3}{2},-\frac{x^2}{\nu}\right)}
{\sqrt{\pi \nu}\, \Gamma\left(\frac{\nu}{2}\right)}\ee
for $F_{1,2}$ hypergeometric function.

\subsection{Absolute central moments method}
For method of moments we will use absolute central moments: $E[|x-\mu|^p]$ for not necessarily integer power $p\in \mathbb{R}^+$. Using Mathematica there was calculated moment formula as the below integral, finite for $p<\nu$:
\be M_{\nu p}=\sqrt[\leftroot{5}p]{\int_{-\infty}^\infty |x|^p \rho_{01\nu}(x) dx}=
\sqrt[\leftroot{5}p]
{\frac{\nu^{p/2}\Gamma\left(\frac{p+1}{2}\right)\Gamma\left(\frac{\nu-p}{2}\right)}
{\sqrt{\pi}\, \Gamma(\nu/2)}  }\label{mom}
\ee
Having a $\{x_t \}_{t=1..T}$ data sample, fixing $\nu$ and using some $\mu$ estimator e.g. approximate $\hat{\mu}=T^{-1} \sum_t x_t$ as just mean, the above formula gives simple estimator of scale parameter $\sigma$:
\be \hat{\sigma} =  \frac{\sqrt[\leftroot{5}p]{T^{-1} \sum_t |x_t-\hat{\mu}|^p}}{M_{\nu p}} \label{sigmaest}\ee
The used $p$ has to be in $(0,\nu)$ range, where the possibility to use non-integer $p$ might be crucial for the $p<\nu$ requirement.

Additionally, using various $p$ for such $\sigma$ estimation has various uncertainty depending on $\nu$, as shown in Fig. \ref{est} - suggesting to optimize $p$ e.g. based on the used $\nu$ range, or even modify $p$ dynamically. For large $\nu$ the optimal $p$ is close to $p=2$ variance estimation, standard for $\nu\to\infty$ Gauss distribution limit. For small $\nu$ the optimal $p$ is $\approx \nu/6$.\\

To estimate $\nu$, a natural direct way is to divide such averages for two different powers $p_1,p_2$, removing $\sigma$ dependence:
\be \frac{M_{\nu p_1}}{M_{\nu p_2}} \approx
\frac{\sqrt[\leftroot{7}p_1]{N^{-1} \sum_i |x_i-\hat{\mu}|^{p_1}}}
{\sqrt[\leftroot{7}p_2]{N^{-1} \sum_i |x_i-\hat{\mu}|^{p_2}}}\label{nuest}\ee
Choosing some $p_1\neq p_2$, the $M_{\nu p_1}/M_{\nu p_2}$ is monotonous with $\nu$ (examples in Fig. \ref{est}), we can e.g. put its behavior into a table and interpolate based on the averages to estimate $\nu$, e.g. done as \verb"find"$\nu$ in the code in Fig. \ref{djiastud}.

However, analogously to $1/(n-1)$ standard adjustment in variance estimator,  (\ref{nuest}) estimation seems biased - needs adjustment by calculating its expected value, preferably with an explicit formula (yet to be found).  In Fig. \ref{djiastud} such slight adjustment was made by just adding (tuned) 0.9 to found $\nu$.\\

Alternatively, as used for stable distribution~(\cite{est,stablebook}), we can use $\nu>p,-p$ opposite powers, multiplying moments instead:
$$E\left[|X|^p\right] E\left[|X|^{-p}\right]=\frac{\Gamma\left(\frac{p+1}{2}\right)\Gamma\left(\frac{\nu-p}{2}\right)}
{\sqrt{\pi}\, \Gamma(\nu/2)}
\frac{\Gamma\left(\frac{-p+1}{2}\right)\Gamma\left(\frac{\nu+p}{2}\right)}
{\sqrt{\pi}\, \Gamma(\nu/2)} $$
However, it lead to inferior log-likelihood for DJIA.

\begin{figure}[t!]
    \centering
        \includegraphics[width=9cm]{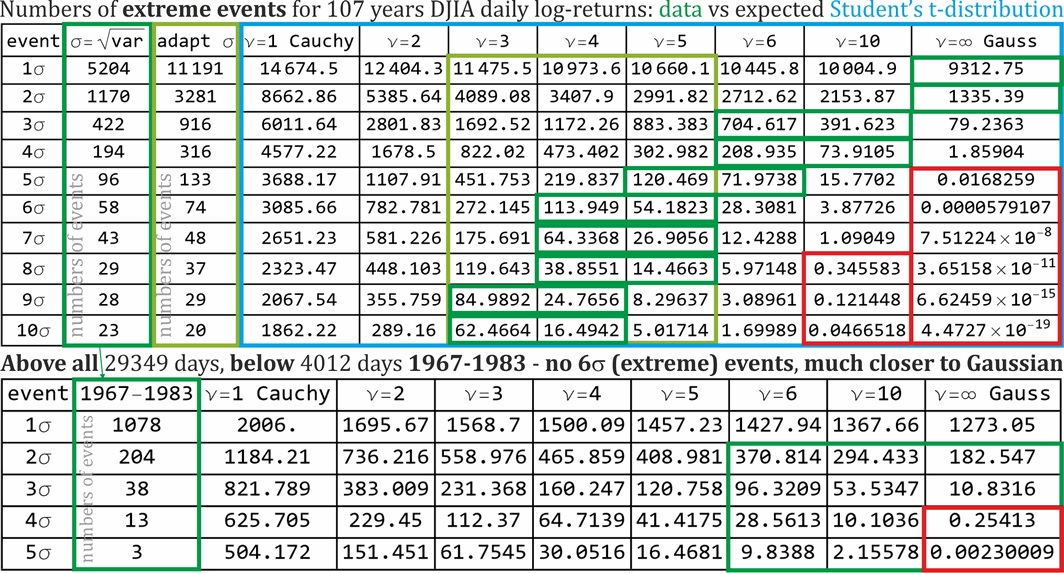}
        \caption{The actual and expected numbers of events $|X-\mu|>k\sigma$: for $k=1,\ldots,10$, complete time series of 29349 values 1900-2007 (top) and restricted to 4012 values 1967-1983 (bottom). The marked green second column are numbers of values in the data, on the right there are expected numbers of events (length $\times$ probability) for Student's t-distribution for various $\nu$. In the top table we see large numbers of extreme events, after using adaptive $\sigma$ close to $\nu\in(3,5)$ Student's t-distribution. In contrast, the 1967-1983 range, suggested by $\nu$ evolution in Fig. \ref{djiastud}, has much lower $\nu\sim 10$ probability of extreme events - suggesting more stable market.  Fig. \ref{nu} shows more detailed $\nu$ evolutions, what might be helpful with localizing, understand the crucial mechanisms, and maybe exploiting them to make the market more stable. }
        \label{table}
\end{figure}

\begin{figure}[t!]
    \centering
        \includegraphics[width=9cm]{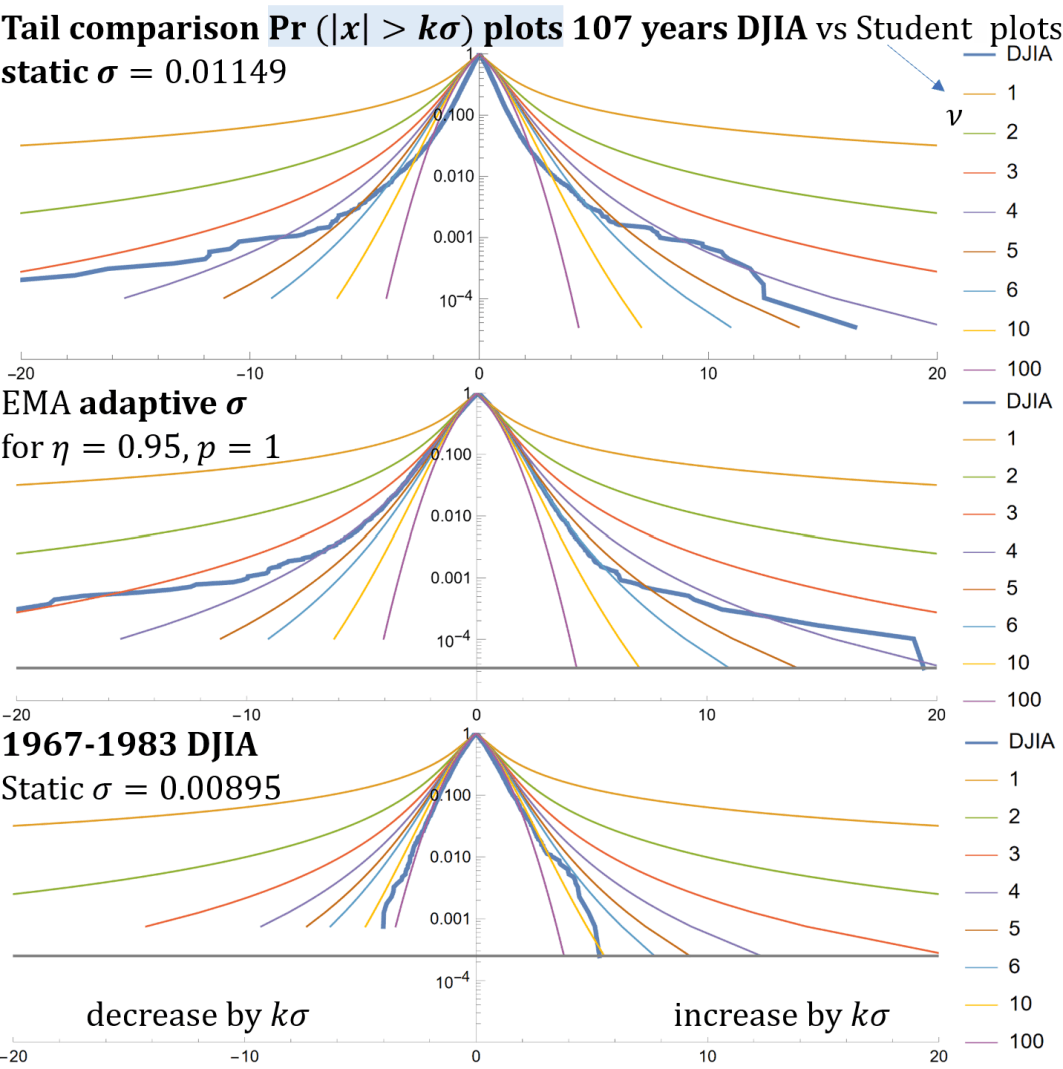}
        \caption{Visualization for Fig. \ref{table} table: probability of exceeding $k\sigma$ toward left (negative) and right (positive): based on DJIA data (bold blue), and its comparison with of Student't-distribution for various $\nu$ degrees of freedom (thin color lines). \textbf{Top}: static $\sigma=0.001149$ for the entire 107 year period, we can observe the central behavior is nearly linear in logarithmic scale as in Laplace distribution. \textbf{Center}: adaptive $\sigma$ makes central behavior closer to Student's t-distribution, but tails corresponding to various $\nu$ between 3 and 6, with visible asymmetry between left and right tails. \textbf{Bottom}: as noticed, 1967-1983 DJIA had nearly Gaussian distribution, what we can see in bottom plot for data restricted to this period, providing good agreement already for static $\sigma$, and with tails for $\nu$ between 10 and 100.}
       \label{tail}
\end{figure}

\subsection{Moving central moments estimators}
Above methods of moments can be easily adapted for moving estimator by just replacing averages with exponential moving averages - uniform weights with exponentially weakening.

For the center $\mu$ we can use just a basic adaptation below - it is optimal only for the Gaussian case ($\nu\to \infty$), hence generally it could be slightly improved. However, for the discussed data the gains were already nearly negligible.
\be \mu_{t+1} = \mu_{t} + \eta_1 (x_t -\mu_t ) \ee

The most crucial is $\sigma$ scale parameter adaptive estimation, as e.g. in ARCH family but in more agnostic way, here using (\ref{sigmaest}) formula for a chosen $p\in (0,2)$ power ($p<\min_t(\nu_t)$), this time with (central absolute) moments evolving in time:
\be m_{p,t+1} = m_{p,t} + \eta_2 (|x_t-\mu_t|^p-m_{p,t}) \ee

Finally, for $\nu$ degrees of freedom estimation we can use (\ref{nuest}) formula for analogously updated moments for some 2 different powers $p_1, p_2$ and some $\eta_3$ learning rate. 

Figure \ref{djiastud} contains used Mathematica code for adaptation of all 3 parameters, with their evolution for DJIA time series. Manual tuning has lead to 3 different learning rates there: $\eta_1=0.003, \eta_2=0.05, \eta_3=0.005$ for correspondingly $\mu,\sigma,\nu$ (much faster for scale parameter $\sigma$).

\begin{figure*}[t!]
    \centering
        \includegraphics[width=19cm]{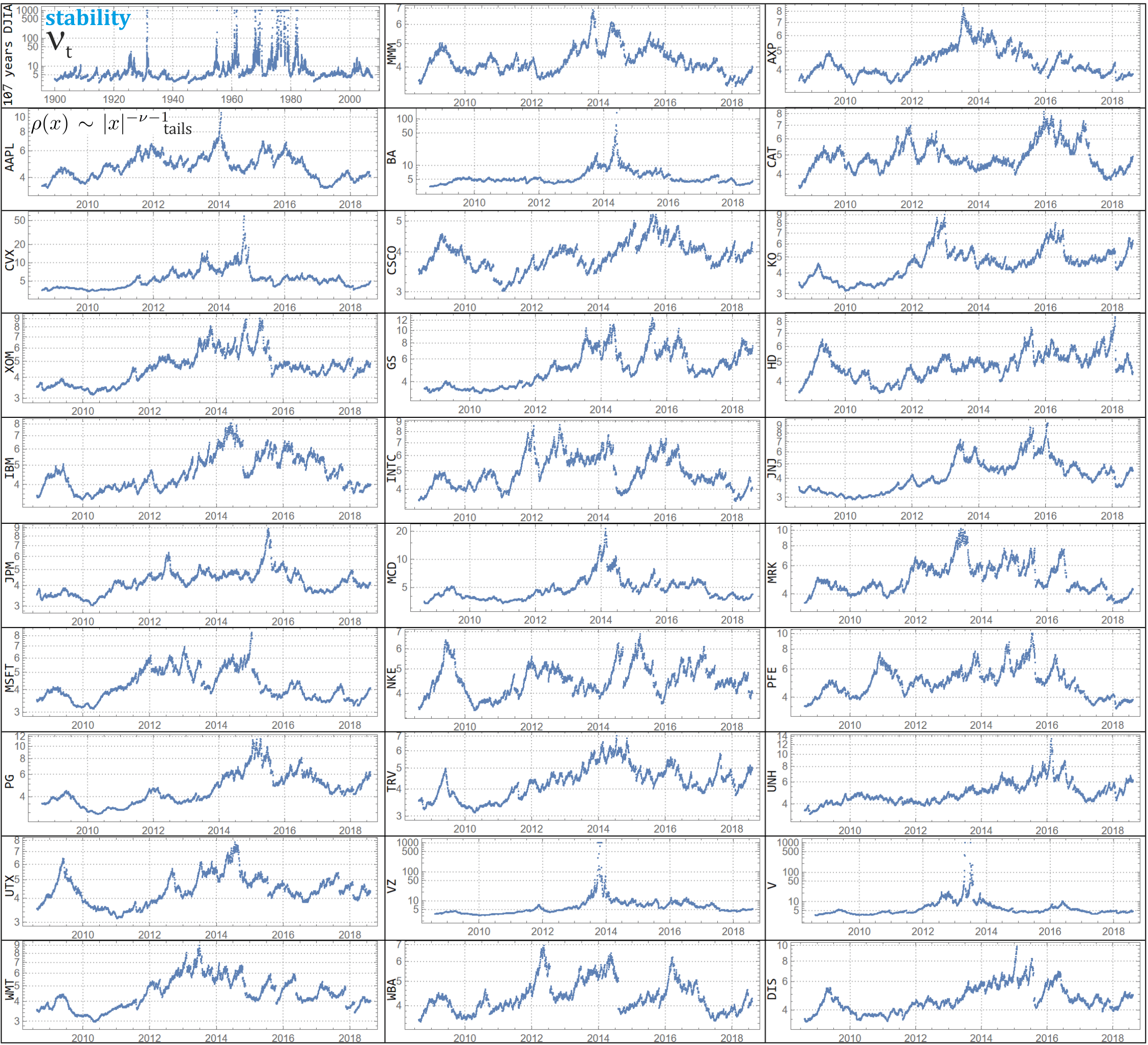}
        \caption{Evolution of $\nu$ parameter for all 1+29 cases with $p_1=1$, $p_2=1/2$ powers and $\eta_3=0.005$ learning rate. It describes tail shape $\rho(x)~\sim |x|^{-\nu-1}$, probability of extreme events - potentially catastrophic, which might destabilize the market, suggesting "stability" interpretation complementing standard "volatility" evaluation. Comparing the above evolutions with various historical events/factors might allow to understand and exploit them to improve market stability.   }
        \label{nu}
\end{figure*}

Figure \ref{eval} shows evaluation using fixed $\mu=0$ center and various fixed $\nu$ for single MLE $\sigma$ parameter, or $\sigma$ adapted using (\ref{sigmaest}) estimation with $p=1$ power and $\eta_2=0.05$ learning rate - e.g. leading to log-likelihood worse only by $\approx 0.004$ than for optimized evolution of all 3 parameters in Fig. \ref{djiastud}. The $\nu$ estimator needs adjustment - here done by just adding tuned parameter, hopefully to be improved, automatized in future.

The $\nu$ evolution, unavailable in standard ARMA-ARCH approaches, evaluates local tail shapes, probability of potentially destabilizing extreme events - suggesting to call it \textbf{stability}, complementing popular \textbf{volatility} evaluation similar to $\sigma$. Figures \ref{table}, \ref{tail} check that indeed 1967-1983 range suggested in Fig. \ref{djiastud} has much thinner tails. Figure \ref{nu} shows $\nu$ evolution for all the companies ($\eta_3=0.005$) - such analysis might help to localize and understand stability influencing factors/mechanisms, which hopefully could be applied in future to reduce probability of potentially catastrophic extreme events.

\subsection{Log absolute moment estimation}
Alternative approach considered e.g. for stable distribution~(\cite{est,stablebook}) is estimation from moments of logarithm of absolute value. We start with transformation of the original variable $X$ to $Y=\ln(|X|)$. 

Transforming its moment generating function:
$$ E\left[\exp(pY)\right]=\sum_{k=0}^\infty E\left[Y^k\right] \frac{p^k}{k!} =E\left[|X|^p\right]=(M_{\nu p} \sigma)^p $$
which allows to express these moments as 
\be E[Y^k]=\frac{d^k}{dp^k} (M_{\nu p} \sigma)^p\Big|_{p=0}\ee
Calculating it for $k=1,2$ using (\ref{mom}) we can get:
$$ E[Y^2]-E[Y]^2=E[(Y-E[Y])^2]=\frac{\pi^2 + \textrm{PolyGamma}(1,\nu/2)}{8} $$
Allowing to estimate $\nu$ from moments, also in adaptive way by their EMA update. However, for the discussed data, such moving $\nu$ estimator has led to slightly worse log-likelihood. 

\begin{figure*}[t!]
    \centering
        \includegraphics[width=19cm]{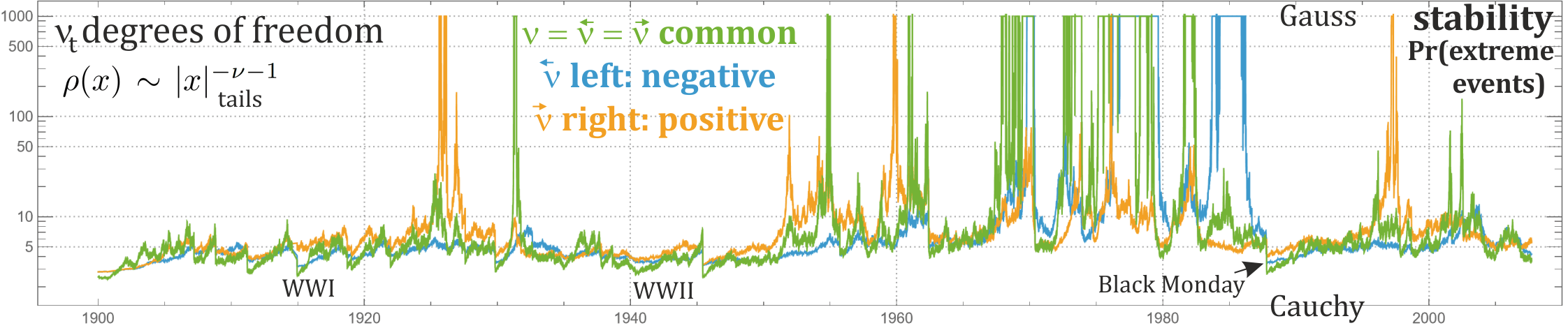}
        \caption{As Fig. \ref{tail} suggested tail asymmetry, there was performed its analysis: adaptive estimation of $\nu$ degrees of freedom for DJIA as previously for the entire distribution in \ref{djiastud} (green), and the same separately: $\overleftarrow{\nu}$ for $x_t < \mu_t$ left tail of negative values (blue) and $\overrightarrow{\nu}$ for $x_t > \mu_t$ right tail of positive values (orange). Interestingly, we can observe regions like 1983-87 with only single heavy tail. Analogous separate estimation of left/right  $\sigma$ was more noisy and only worsened log-likelihood, so it is not presented. }
        \label{asymmetric}
\end{figure*}

\section{Including asymmetry}
As especially the tails e.g. in Fig. \ref{tail} analysis are clearly asymmetric, it might be also valuable to include it. There are two classical approaches to asymmetrize Student-t distribution: \emph{noncentral t-distribution}~\cite{noncentral} with PDF for $\nu,\delta$ parameters:
$$ \frac{e^{-\frac{\delta ^2}{2}} 2^{\nu } \nu ^{\frac{\nu }{2}+1} \Gamma \left(\frac{\nu +1}{2}\right) }{\pi }H_{-\nu -1}\left(\frac{-x \delta }{\sqrt{2} \sqrt{x^2+\nu }}\right) \left(\nu +x^2\right)^{-\frac{\nu +1}{2}}$$
and \emph{skewed generalized t-distribution}~\cite{skewed} with PDF:
$$ \frac{p}{2 v \sigma  q^{\frac{1}{p}} B(\frac{1}{p},q) \left(1 + \frac{| x-\mu + m |^p}{q (v \sigma)^p (1 + \lambda \text{sgn}(x-\mu + m))^p}\right)^{\frac{1}{p}+q}} $$

However, they are much more complicated, especially if wanting to search for adaptive estimation, what might be worth to consider in the future.

Instead, for simplicity let us consider asymmetry by just gluing in $\mu$ two standard Student t densities of separate parameters: $\overleftarrow{\sigma},\overleftarrow{\nu}$ describing the part on the left $x<\mu$, and $\overrightarrow{\sigma},\overrightarrow{\nu}$ describing the part on the right $x>\mu$. While it is not necessary, wanting continuous PDF we need to enforce that both parts have the same value in $\mu$, what through linear scaling leads to PDF:

\begin{figure}[t!]
    \centering
        \includegraphics[width=9cm]{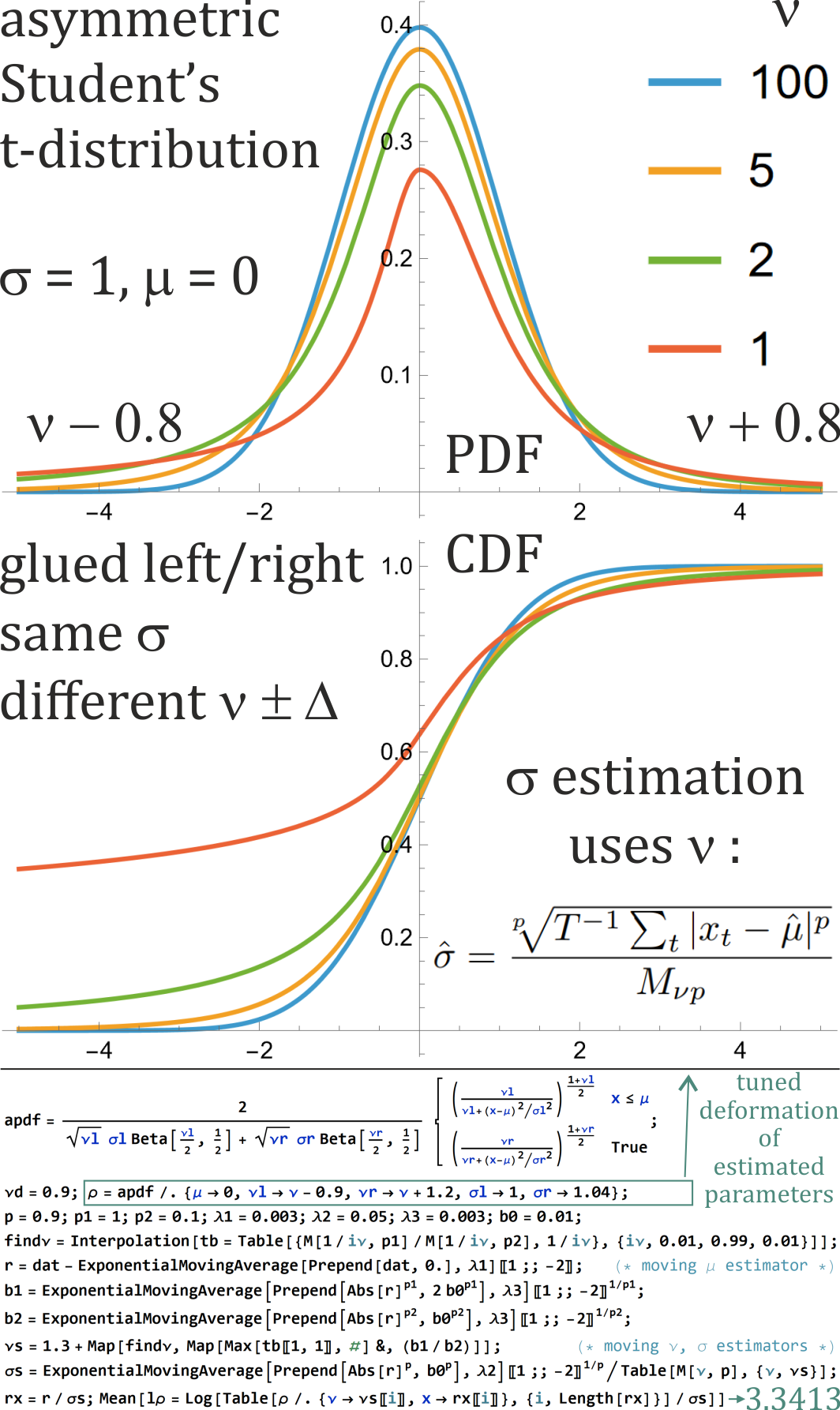}
        \caption{Top: proposed asymmetric Student's t-distribution (\ref{ast}) as just glued standard left/right of different $\sigma,\nu$, rescaled for continuity. Here of the same $\sigma=1$, but optimized for DJIA series: with left $\nu$ lower by 0.8, right higher by 0.8 - heavier tails for price drops, while $\nu$ is used for estimation of $\sigma$. Bottom: the used source for further deformed parameters to maximize log-likelihood on DJIA series: with fixed $\nu$ differences, additionally slightly deforming $\sigma$, allowing to improve log-likelihood from the original 3.3389 to 3.3413, which  density and parameter evolution is presented in Fig. \ref{fin}.}
       \label{astudent}
\end{figure}
\begin{figure*}[t!]
    \centering
        \includegraphics[width=19cm]{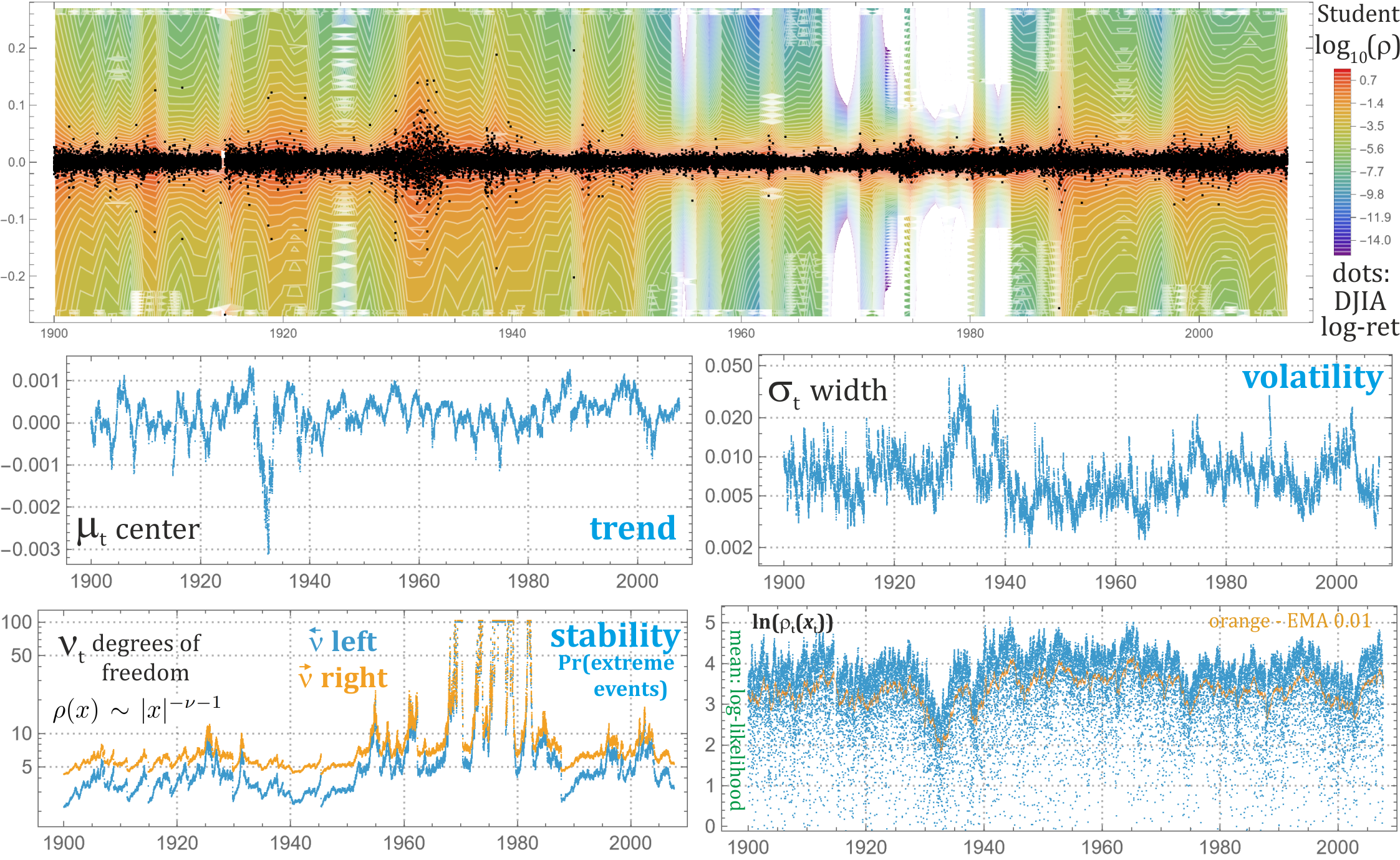}
        \caption{Top: DJIA 107 years log-returns (black points) and base-10 logarithms of densities from the best found evolving asymmetric Student's t-distribution model using source from Fig. \ref{astudent}. We can observe that price drops use slightly heavier tails ($\nu$ smaller by 2.1). Bottom: used evolution of their parameters. }
        \label{fin}
\end{figure*}

\be\rho_{\mu\overleftarrow{\sigma}\overleftarrow{\nu}\overrightarrow{\sigma}\overrightarrow{\nu}}(x)=
\frac{2 \left\{
\begin{array}{cc}
 \left(1+\frac{(x-\mu )^2}{\overleftarrow{\sigma}^2\,\overleftarrow{\nu} }\right)^{-\frac{\overleftarrow{\nu}+1}{2}}  \text{ if }x\leq \mu  \\
 \left(1+\frac{(x-\mu )^2}{\overrightarrow{\sigma}^2\,\overrightarrow{\nu}}\right)^{-\frac{\overrightarrow{\nu}+1}{2}} \text{ if }x> \mu \\
\end{array}
\right.
 }{\sigma_l \sqrt{\overleftarrow{\nu}}  B\left(\frac{\overleftarrow{\nu}}{2},\frac{1}{2}\right)+ \sigma_r\sqrt{\overrightarrow{\nu}}  B\left(\frac{\overrightarrow{\nu}}{2},\frac{1}{2}\right)}\label{ast}\ee 
 
It has advantage that we can use adaptive estimation of $\sigma,\nu$ exactly as previously, just separate for values below/above current $\mu$: we maintain two copies of $\sigma,\nu$, and update one of them based on $\textrm{sgn}(x-\mu)$. 

However, tests on DJIA has lead to $\approx 0.01$ worse log-likelihood for separate left/right adaptation of both $\overleftarrow{\nu},\overrightarrow{\nu}$ and $\overleftarrow{\sigma},\overrightarrow{\sigma}$. It has returned to the original log-likelihood for separate $\overleftarrow{\nu},\overrightarrow{\nu}$ adaptation and common $\sigma=\overleftarrow{\sigma}=\overrightarrow{\sigma}$. So while $\overleftarrow{\nu},\overrightarrow{\nu}$ can be used to describe shape of separate left/right tail, for $\sigma$ scale parameter it seems better to use a common one. Also we could search for a more sophisticated adaptive estimation, or maybe use the noncentral or skewed t-distribution.

We can use such evolving especially $\nu$ to evaluate the market: e.g. to estimate probability of extreme events separately toward left and right, also use them as local parameters for various models. Figure. \ref{asymmetric} shows its evolution for 107 years of DJIA. 

Later improvement of log-likelihood was reached by estimating $\sigma,\nu$ together, but then using deformed asymmetric Student's t-distribution (\ref{ast}) of tuned parameters, like in Fig. \ref{astudent}. As usually price drop left tail is heavier: has lower $\nu$, and $\overrightarrow{\nu}-\overleftarrow{\nu} \sim 2$ in Fig. \ref{asymmetric} here. Fixing this difference like in Fig. \ref{astudent} has finally allowed to improve log-likelihood from 3.3389 to 3.3406, by using common $\sigma=\overleftarrow{\sigma}=\overrightarrow{\sigma}$ estimated using $\nu$ online updated as previously, but with asymmetric $\overleftarrow{\nu}=\nu+\Delta_\nu,\overrightarrow{\nu}=\nu-\Delta_\nu$ for optimized $\Delta_\nu=0.8$. Its further tuning, also slightly deforming $\sigma$ (suggested by left/right probability difference due to rescaling for (\ref{ast}) continuity), has allowed to increase log-likelihood to 3.3413 using source shown in Fig. \ref{astudent}, its  density and parameter evolution is presented in Fig. \ref{fin}.
\section{Conclusions and further work}
This article introduces looking novel extensions of method of moments - both to absolute central moments with not necessarily natural powers (crucial to work with low $\nu$), but more importantly as EMA moving estimators - for parameters evolving in time, also asymmetrically for left/right tails. Beside better log-likelihood evaluation, it provides evolution of these crucial parameters like in Fig. \ref{djiastud}, \ref{nu}, \ref{asymmetric} - including $\nu$ degrees of freedom evaluating probability of extreme events, which dependency understanding might allow to introduce some market stabilizing mechanisms. For example it suggests search for mechanisms of drastic increase of $\nu$ especially in 1967-1983 period for DJIA, confirmed in Fig. \ref{table}, \ref{tail}.

This is a general approach which might be worth taking also to other distributions like alpha-stable, and larger models. Also it is worth combining with other especially adaptive models, like online linear regression and HCR (hierarchical correlation reconstruction) - what is planned to be done in further versions of this article.

Examples of plans for further work:
\begin{itemize}
  \item Improve estimators from moments - especially of $\nu$.
  \item Improve evolution for asymmetric cases, e.g. better adaptive estimation of the proposed glued two Student t-distributions, or noncentral, skewed variants.
  \item Add further modelling, like dependence from other stocks, macronomical data, e.g. with adaptive linear regression~\cite{regression}, and HCR~\cite{hcr} to include subtle dependencies.
  \item Find various approaches for moving estimators of various distributions, e.g. with gradient ascend approaches, maybe also including 2nd order information like in \cite{OGR}.
  \item The discussed approach has many hyperparameters like learning rates -  often universal for similar data types. It might be valuable to automatically optimize them, adapt through evolution.
  \item Understand mechanisms/dependencies affecting $\nu$ evolution, also separate for left/right tail, and hopefully exploit them e.g. to improve marked stability.
  \item Test discussed approaches for different application like data compression, where log-likelihood improvement translates into nit/symbol savings.
  \item Applications for online estimation of Hurst exponent, closely related with heavy tail distributions like Student's t~\cite{sest} or stable~\cite{est}.
\end{itemize}

\bibliographystyle{IEEEtran}
\bibliography{cites1}

\begin{thebibliography}{10}
\providecommand{\url}[1]{#1}
\csname url@samestyle\endcsname
\providecommand{\newblock}{\relax}
\providecommand{\bibinfo}[2]{#2}
\providecommand{\BIBentrySTDinterwordspacing}{\spaceskip=0pt\relax}
\providecommand{\BIBentryALTinterwordstretchfactor}{4}
\providecommand{\BIBentryALTinterwordspacing}{\spaceskip=\fontdimen2\font plus
\BIBentryALTinterwordstretchfactor\fontdimen3\font minus
  \fontdimen4\font\relax}
\providecommand{\BIBforeignlanguage}[2]{{%
\expandafter\ifx\csname l@#1\endcsname\relax
\typeout{** WARNING: IEEEtran.bst: No hyphenation pattern has been}%
\typeout{** loaded for the language `#1'. Using the pattern for}%
\typeout{** the default language instead.}%
\else
\language=\csname l@#1\endcsname
\fi
#2}}
\providecommand{\BIBdecl}{\relax}
\BIBdecl

\bibitem{adaptive}
J.~Duda, ``Adaptive exponential power distribution with moving estimator for
  nonstationary time series,'' \emph{arXiv preprint arXiv:2003.02149}, 2020.

\bibitem{garch}
T.~Bollerslev, ``Generalized autoregressive conditional heteroskedasticity,''
  \emph{Journal of econometrics}, vol.~31, no.~3, pp. 307--327, 1986.

\bibitem{OPCA}
H.~Cardot and D.~Degras, ``Online principal component analysis in high
  dimension: Which algorithm to choose?'' \emph{International Statistical
  Review}, vol.~86, no.~1, pp. 29--50, 2018.

\bibitem{regression}
J.~Duda, ``Parametric context adaptive laplace distribution for multimedia
  compression,'' \emph{arXiv preprint arXiv:1906.03238}, 2019.

\bibitem{hcr}
------, ``Exploiting statistical dependencies of time series with hierarchical
  correlation reconstruction,'' \emph{arXiv preprint arXiv:1807.04119}, 2018.

\bibitem{student}
F.~R. Helmert, ``{\"U}ber die berechnung des wahrscheinlichen fehlers aus einer
  endlichen anzahl wahrer beobachtungsfehler,'' \emph{Z. Math. U. Physik},
  vol.~20, no. 1875, pp. 300--303, 1875.

\bibitem{student1}
Student, ``The probable error of a mean,'' \emph{Biometrika}, vol.~6, no.~1,
  pp. 1--25, 1908.

\bibitem{est}
B.~Basu and V.~Pakrashi, ``Parameter estimates of alpha-stable distribution and
  hurst coefficients,'' \emph{Journal of Environmental Engineering and
  Science}, vol.~13, no.~3, pp. 53--72, 2018.

\bibitem{stablebook}
C.~L. Nikias and M.~Shao, \emph{Signal processing with alpha-stable
  distributions and applications}.\hskip 1em plus 0.5em minus 0.4em\relax
  Wiley-Interscience, 1995.

\bibitem{noncentral}
N.~Johnson and B.~Welch, ``Applications of the non-central t-distribution,''
  \emph{Biometrika}, vol.~31, no. 3/4, pp. 362--389, 1940.

\bibitem{skewed}
P.~Theodossiou, ``Financial data and the skewed generalized t distribution,''
  \emph{Management science}, vol.~44, no. 12-part-1, pp. 1650--1661, 1998.

\bibitem{OGR}
J.~Duda, ``{Improving SGD convergence by tracing multiple promising directions
  and estimating distance to minimum},'' \emph{arXiv preprint
  arXiv:1901.11457}, 2019.

\bibitem{sest}
K.~E. Bassler, G.~H. Gunaratne, and J.~L. McCauley, ``Markov processes, hurst
  exponents, and nonlinear diffusion equations: With application to finance,''
  \emph{Physica A: Statistical Mechanics and its Applications}, vol. 369,
  no.~2, pp. 343--353, 2006.

\end{thebibliography}
\end{document}